\documentclass[9pt,twocolumn,twoside]{osajnl}
\usepackage{multirow}

\journal{optica} 

\setboolean{shortarticle}{true}

\setboolean{displaycopyright}{true}

\begin{document}

\author[1]{John M. Robinson}
\author[1,5]{Eric Oelker}
\author[1]{William R. Milner}
\author[1]{Dhruv Kedar}
\author[1,3]{Wei Zhang}
\author[2]{Thomas Legero}
\author[2,4]{Dan G Matei}
\author[2]{Sebastian Häfner}
\author[2]{Fritz Riehle}
\author[2]{Uwe Sterr}
\author[1]{Jun Ye}

\affil[*]{Corresponding author: john.robinson@colorado.edu}




\setboolean{displaycopyright}{true}

\title{Thermal noise and mechanical loss of SiO$_2$/Ta$_2$O$_5$ optical coatings at cryogenic temperatures}

\affil[1]{JILA, NIST and University of Colorado, 440 UCB, Boulder, Colorado 80309, USA}
\affil[2]{Physikalisch-Technische Bundesanstalt, Bundesallee 100, 38116 Braunschweig, Germany}
\affil[3]{Current address : Jet Propulsion Laboratory, 4800 Oak Grove Drive Pasadena, CA 91109  
}

\affil[4]{Current address : Horia Hulubei National Institute of Physics and Nuclear Engineering, Reactorului 30, 077125 Magurele, Romania 
}

\affil[5]{Current address : School of Physics and Astronomy, SUPA, University of Glasgow, Glasgow G12 8QQ, UK}


\date{\today}

\begin{abstract}
Mechanical loss of dielectric mirror coatings sets fundamental limits for both gravitational wave detectors and cavity-stabilized optical local oscillators for atomic clocks. 
Two approaches are used to determine the mechanical loss: ringdown measurements of the coating quality factor and direct measurement of the coating thermal noise. 
Here we report a systematic study of the mirror thermal noise from  room temperature to 4 K by operating reference cavities at these temperatures. 
The directly measured thermal noise is used to extract the corresponding mechanical loss for SiO$_2$/Ta$_2$O$_5$ coatings, which are compared with previously reported values.
\end{abstract}

\maketitle

\thispagestyle{empty}
Low mechanical loss optical coatings play a critical role in optical reference cavities \cite{kessler_2012, Matei_PRL, Zhang_PRL_2017, Robinson_2019}, gravitational wave detectors \cite{Harry:06}, and cavity optomechanics \cite{Cripe_2019}.
Mechanical dissipation in the coating often dictates the amplitude of length fluctuations.
These fluctuations are referred to as Brownian coating thermal noise (CTN), and can limit the precision of interferometric measurements.
Thus, there has been considerable experimental effort to characterize the level of CTN present in current optical coatings.

The power spectral density (PSD) of thermal noise induced displacement fluctuations is proportional to temperature.
For this reason, several gravitational wave detectors plan to operate at cryogenic temperatures, including KAGRA at 20 K \cite{Akutsu_2019}, LIGO Voyager at 124 K \cite{shapiro_ligovoyager, LIGO_Cosmic_Explorer}, and the Einstein Telescope at 10 K~\cite{ET_design}.
It is also clear that cryogenic optical cavities provide the best laser stability for optical atomic clocks~\cite{oelker_2019,Matei_PRL}.
Improving the performance of these inferferometers relies upon characterizing and improving the CTN at cryogenic temperatures.

There are two distinct methods for determining the Brownian CTN of an optical coating.
The first is referred to as the "mechanical ringdown approach". 
This involves measuring the mechanical quality factor, Poisson ratio, and Young's modulus for each coating material. 
These mechanical properties for the coating and substrate are then used to calculate the CTN for a given mirror.
This method of calculating the CTN may not account for all multilayer phenomena in the coating. 
The second method, referred to as "direct measurements", determines the CTN of the coating by measuring the frequency stability of optical cavities.
This approach has the challenge of extracting a coating loss angle from a single measurement over a broad frequency range, which requires input on other coating and substrate properties.
Due to the challenges in each approach, it is vital that both are undertaken as independent and complementary research efforts.
In this Letter, we present direct CTN measurements of SiO$_2$/Ta$_2$O$_5$ using the most stable cavities in operation today at cryogenic temperatures up to room temperature.

Highly reflective optical coatings are fabricated by stacking alternating layers of high and low refractive index material.
The most common choices are SiO$_2$ and Ta$_2$O$_5$ for the low and high refractive index, respectively.
Mechanical ringdown measurements have been used to characterize both the coating materials and multilayer coatings at cryogenic temperatures.
Individual thin films of SiO$_2$ and Ta$_2$O$_5$ have shown a peak in the mechanical loss angle at cryogenic temperatures~\cite{Glasgow_PRL_2019, Martin_2010}.
A multilayer coating of SiO$_2$/Ti:Ta$_2$O$_5$ was measured to have a loss peak, reaching approximately $1.0\times 10^{-3}$ at $T = 20$ K \cite{Granata_CryoTa2O5TiO2}.
Ringdown measurements of an undoped SiO$_2$/Ta$_2$O$_5$ coating showed a largely temperature independent mechanical loss angle of $4~\times~10^{-4}$ at cryogenic temperatures \cite{Yamamoto_2006}.
The possibility of increased mechanical loss at cryogenic temperatures has motivated research in developing new optical coatings~\cite{Glasgow_PRL_2019, Cole_2013, Cole_2016_Optica}.
We are presenting our direct CTN measurements versus temperature to complement the growing body of CTN research.

We perform direct measurement of CTN of two independent multilayer coatings composed of SiO$_2$/Ta$_2$O$_5$. By employing these coatings in ultrastable rigid crystalline silicon (c-Si) Fabry-Perot cavities, we characterize the CTN around 4 K, 16 K, and 124 K.
These particular temperatures are informative for future gravitational wave detectors.
We also show an upper bound for a third SiO$_2$/Ta$_2$O$_5$ coating at room temperature, measured using an ultra-low expansion (ULE) cavity.

We study the mechanical loss of the HR coatings by measuring the noise of three independent Fabry-Perot cavities~\cite{Matei_PRL, Robinson_2019, Swallows_2012}.
All three cavities use Ion-Beam-Sputtered optical coatings made by Advanced Thin Films.
The coatings for the cryogenic cavities have 21 layers of SiO$_2$ and 20 layers of Ta$_2$O$_5$, while the room temperature system has 18 SiO$_2$ and 19 Ta$_2$O$_5$ layers respectively.
The coatings were deposited at room temperature, and then annealed at a temperature of approximately $480 ^{\circ}$C.
The physical details of the cavities are shown in table \ref{tab:table_cavities}, where $L$ is the cavity length, $T$ is the cavity temperature, $d$ is the coating thickness, and ROC$_{1,2}$ is the radius of curvature of each mirror, respectively.
We calculate individual contributions to thermal noise for each system according to the analysis in \cite{CoatingBook_martin}.
Here, noise is quantified in terms of fractional frequency PSD, which is defined as 
$S_{y}(f) = \frac{S_{\nu}(f)}{\nu^2}$,
where $S_{\nu}(f)$ is the single-sided PSD of frequency fluctuations and $\nu$ is the optical frequency.
\begin{figure}
    \centering
    \includegraphics[width=7.5cm]{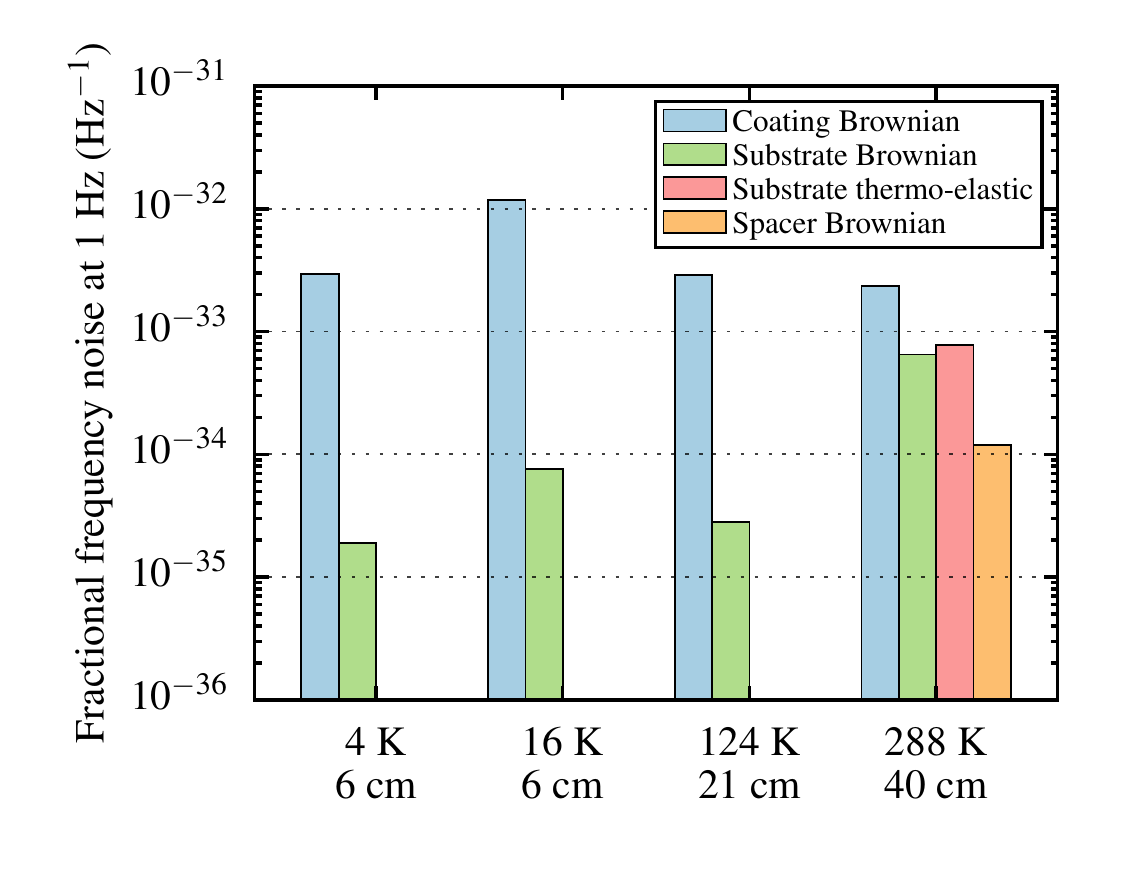}
    \caption{Based on representative numbers from existing literature, we illustrate the major contributions to the thermal noise for three optical cavities at various temperatures and with length ranging from 6 to 40 cm. Fractional noise associated with coating and substrate scales inversely with the cavity length. Substrate thermo-elastic and spacer Brownian noise are below $10^{-36}$ Hz$^{-1}$ for our cryogenic systems.}
    \label{fig:fig1}
\end{figure}
\begin{table}[!htbp]
\centering
\caption{ \bf Optical cavity and coating properties}
\begin{tabular}{cccccc}
\hline
 L~(cm) & T~(K) & d~($\mu$m) & Spacer/Sub. & ROC$_1$/ROC$_2$\\ \hline
 6   & 4,16.2 & 9.2 &  c-Si/c-Si & 1m/1m \\
 21  & 123.3 & 9.2 & c-Si/c-Si & $\infty$/1m \\
 40  & 288 & 3.8 & ULE/FS & $\infty$/1m  \\
\end{tabular}
\label{tab:table_cavities}
\end{table}

The contributions from Brownian motion of the cavity constituents and substrate thermo-elastic noise are shown in Fig.~\ref{fig:fig1}.
We use the material properties in Table~\ref{tab:material_properties} and the equations in \cite{CoatingBook_martin}.
\begin{table}
\centering
\caption{\label{tab:material_properties} \bf Material properties}
\begin{tabular}{cccccc}
 Material  &Y (GPa)  &$\sigma$  & $\phi$  \\ \hline
 c-Si ($\langle 111 \rangle$)  & 187.5~\cite{Brantley_Si}  &0.23~\cite{Brantley_Si}  & $10^{-7}$~\cite{Nawrodt_2008}   \\
 SiO$_2$-Ta$_2$O$_5$   &  91(7) \cite{Crooks_2006}  & 0.19~\cite{Crooks_2006} & [This work]  \\
 Fused silica & 72 ~\cite{musikant_1985}  & 0.17~\cite{musikant_1985} & $1\times10^{-6}$~\cite{Numata_2000} \\
 ULE & 68~\cite{Klocek_handbook} & 0.17~\cite{Klocek_handbook} & $1.6\times 10^{-5}$~\cite{Numata_2004}
\end{tabular}
\end{table}
Substrate thermo-elastic noise scales with the squared power of temperature $T^2$ and the substrate coefficient of thermal expansion (CTE) $\alpha_{sub}^2$, and is thus less than $10^{-36}$~Hz$^{-1}$ at 1 Hz for our cryogenic systems~\cite{CoatingBook_martin}.
Coating thermo-optic noise is $4~\times 10^{-36}$~Hz$^{-1}$ at 1 Hz for the room temperature system and is below $10^{-36}$~Hz$^{-1}$ at 1 Hz for the cryogenic systems.
The relative magnitude of the spacer and substrate contribution for the cryogenic silicon cavities highlights the advantage of using these systems for studying the CTN, provided one operates at a temperature where the silicon CTE is sufficiently small.

\begin{figure}
    \centering
    \includegraphics[width=7.5cm]{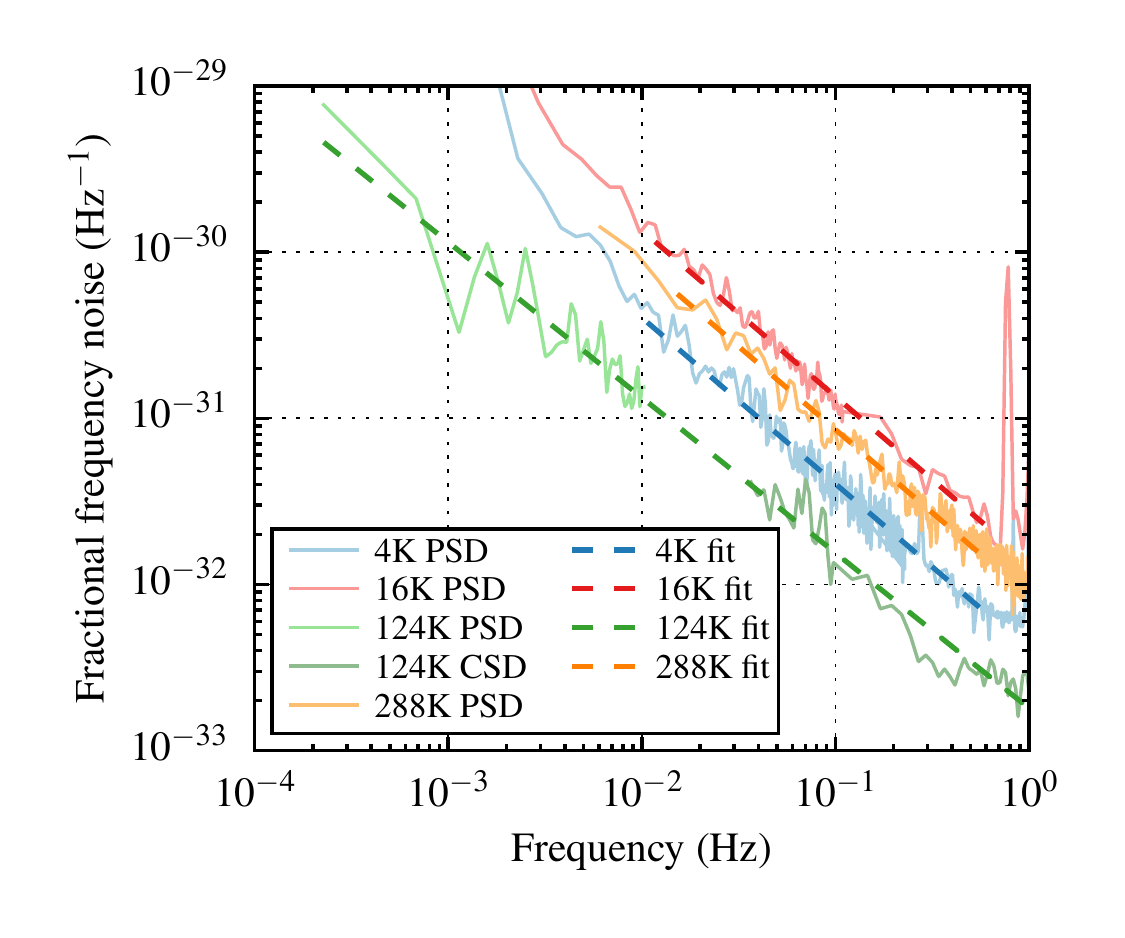}
    \caption{The measured noise PSD at various temperatures (solid curves). The dashed lines are the fits to $S_{y}(f) =  a/f^b$.}
    \label{fig:PSD}
\end{figure}

We first characterize the 124 K system, which has the lowest thermal noise floor as seen in Fig.~\ref{fig:fig1}.
This system is used as the local oscillator for the JILA Sr1 optical lattice clock~\cite{bothwell_metrologia}.
The laser frequency measurement comes from recording the correction signal applied to keep the laser on resonance with the clock transition.
We use three independent data runs, each lasting at least 20,000 seconds.
The PSD is calculated for Fourier frequencies from $3\times 10^{-4}$ to $10^{-2}$ Hz and is shown in Fig.~\ref{fig:PSD} as the green data.
These measurements are restricted to low frequencies, limited by the finite duty cycle of the clock.
We also plot an estimate of the PSD for this system at higher Fourier frequencies based on a cross-spectral density (CSD) measurement from~\cite{oelker_2019}.

The PSD of the 124 K system measured with the optical clock constitutes a direct measurement of its thermal noise level as the noise contribution from the clock itself is negligible~\cite{oelker_2019}.  
We extract the CTN level of the remaining systems by measuring their heterodyne beat against the 124 K system using a lambda-type zero dead time frequency counter, and subtracting it's thermal noise contribution in quadrature.
Since the thermal noise floors are uncorrelated, we subtract the direct measurement in quadrature from each heterodyne beat to obtain the 4 K (blue), 16 K (red), and 288 K (orange) noise PSD.
The uncertainty of the fitted 124 K PSD is propagated when we perform the subtraction.

Several authors have reported a frequency-dependent loss angle for SiO$_2$/Ta$_2$O$_5$ coatings~\cite{Amato_2018, Gras_evans_2018}.
We fit the measured spectra to the functional form of $S_{y}(f) = a / f^b$ in order to allow for any potential frequency dependence.
A deviation from $b = 1$ for the CTN indicates a frequency-dependent loss angle.
The fits to the measured spectra are shown as dashed lines in Fig.~\ref{fig:PSD}. 
We find weak dependencies on frequency as summarized in Fig. 4.

To extract a mechanical loss angle from the measured noise spectrum, we perform a characterization of all other noise sources in the reference cavity systems.
The 124 K system has been thoroughly characterized as described in~\cite{Matei_2016}.
Since that work we have made several improvements to the setup, including active stabilization of the transmitted optical power, active temperature control of the outermost vacuum chamber, and improved optical quality of the optics. 
\begin{figure}
    \centering
    \includegraphics[width=7.55cm]{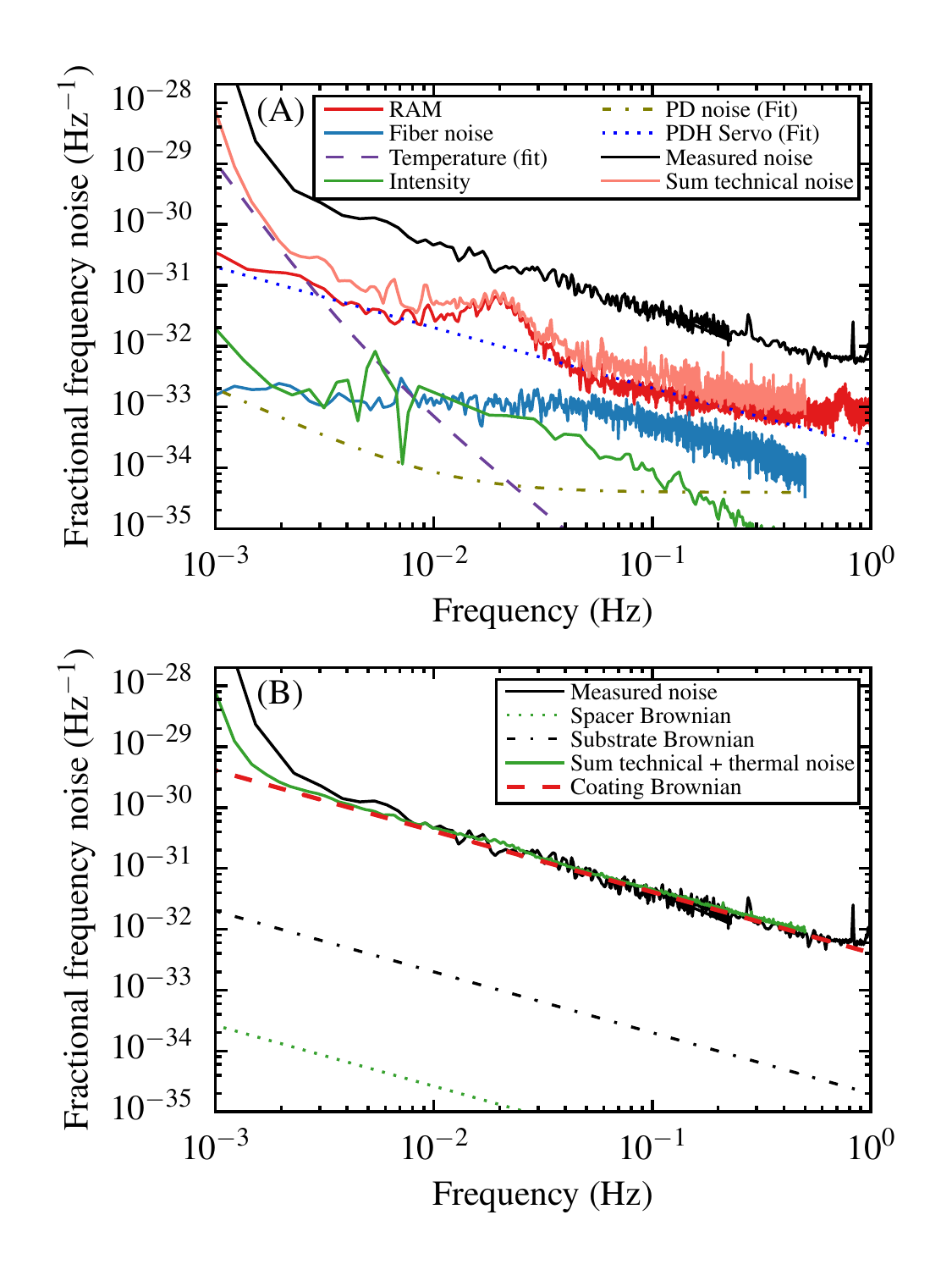}
    \caption{(A) Technical noise contributions for the 4 K system, along with the sum of the technical noise (pink line) and the measured noise (black line). (B) Intrinsic thermal noise for the cavity, along with the sum of technical noise and thermal noise (green) and coating Brownian noise (dashed-red). }
    \label{fig:NoiseBudget}
\end{figure}

The technical noise budget of the 6~cm system is shown in Fig.~\ref{fig:NoiseBudget}A.
We show fits to the measured noise arising from temperature, Pound-Drever-Hall (PDH) servo, and the PDH photodiode (PD).
The dominant technical noise source is residual amplitude modulation (RAM), shown as the red data.
The sum of the technical noise is well below the measured noise.
Figure~\ref{fig:NoiseBudget}B shows the intrinsic thermal noise of the cavity, including Brownian noise from the spacer (dotted), substrate (dash-dot) and coating (dashed).
The measured noise is shown (black) along with the sum of technical noise and thermal noise (green).
The CTN is the dominant noise source from 5 mHz to 0.8 Hz.

The room temperature ULE cavity was previously used as the clock laser for the JILA Sr lattice clock~\cite{Swallows_2012}.
This system has routinely performed at a fractional frequency stability of $1\times 10^{-16}$.
Since we have not developed a thorough noise budget for this system, we report only an upper limit on the loss angle.

With the technical noise on the cryogenic cavities being sufficiently low, we can extract a mechanical loss angle at each respective temperature.
We assume that the mechanical loss in the parallel and perpendicular directions are identical ($\phi_{\parallel} = \phi_{\perp}$)~\cite{FOOTNOTE}.
For the i$^{th}$ mirror, the expression for the fractional frequency PSD arising from coating Brownian noise is then~\cite{Harry_2002}
\begin{equation}
\begin{aligned}
\label{eq:Brownian_coating}
    S_{y}^{i}(f) = & \frac{2 k_B T d}{\pi^{2} f L^2} \frac{1-\sigma_{sub}^2}{\omega_i Y_{sub}^2} \frac{\phi_{c}(f)}{ Y_c (1-\sigma_c^2)(1-\sigma_{sub}^2)}  \times \\ &\left[Y_c^2 (1+\sigma_{sub})^2 (1-2\sigma_{sub})^2 + Y_{sub}^2 (1+\sigma_c)^2 (1-2\sigma_c) \right] 
\end{aligned}
\end{equation}
Here, "c" labels the coating and "sub" labels the substrate. $\sigma_{c(sub)}$ is the Poisson's ratio, $Y_{c(sub)}$ is Young's modulu, $d$ is the coating thickness, $\omega_i$ is the $1/e^2$ beam radius at the $i^{th}$ mirror, $\phi_{c(sub)}$ is the loss angle.
Note that we are allowing for the coating loss angle to have frequency dependence.

\begin{figure}
    \centering
    \includegraphics[width=8.0cm]{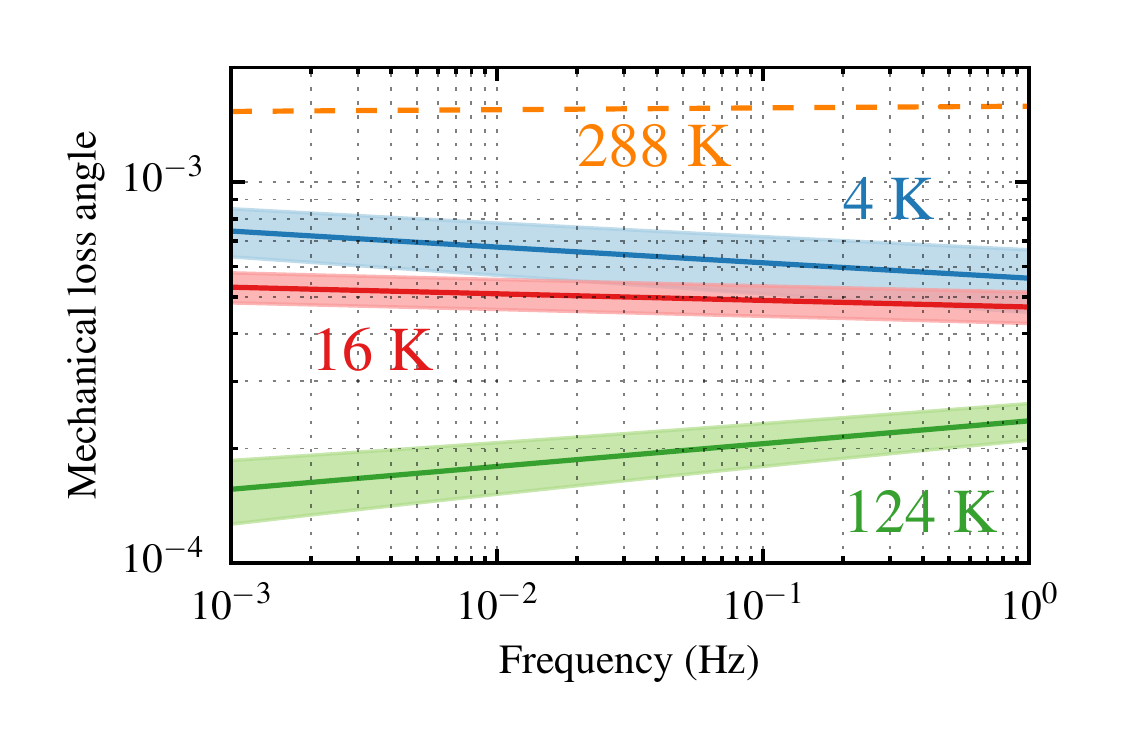}
    \caption{Frequency-dependent mechanical loss angle for the SiO$_2$/Ta$_2$O$_5$ coatings at different operating temperature. The shaded bands indicate the $1\sigma$ uncertainty.}
    \label{fig:freqdependent}
\end{figure}

With the fits from Fig.~\ref{fig:PSD} and Eq.~\ref{eq:Brownian_coating}, we derive $\phi_{c}(f)$ and plot the results in Fig.~\ref{fig:freqdependent}.
The results are
\begin{align}
\phi_{4~K}(f) &= (5.6 \pm 0.9)\times 10^{-4}~\times f^{~-0.05\pm0.01}\\
\phi_{16~K}(f) &= (3.2 \pm 0.3)\times 10^{-4}~\times f^{~-0.11\pm0.02}\\
\phi_{124~K}(f) & =  (2.4 \pm 0.3)\times 10^{-4}~\times f^{~0.06\pm0.02}
\end{align}
The shaded band indicates the 1$\sigma$ uncertainty.
Interestingly, we find that the frequency dependence, although rather weak, has the opposite sign at 4 K and 16 K compared with the 124 K result. 
The 4 and 16~K results use mirrors from a different coating run than the 124~K cavity, and so this difference could be coating-dependent.

\begin{figure}
    \centering
    \includegraphics[width=3.19in]{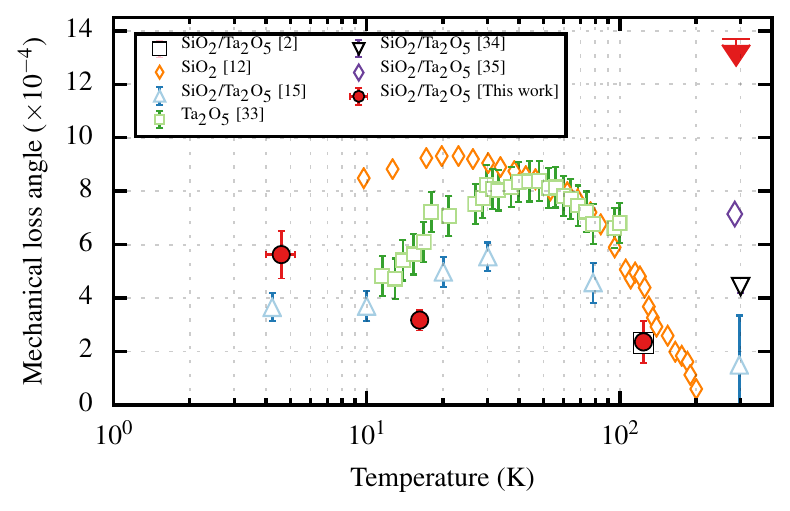}
    \caption{Mechanical loss of the individual coating constituents (SiO$_2$, Ta$_2$O$_5$), and SiO$_2$/Ta$_2$O$_5$ coatings measured by ringdown (orange diamonds, blue triangles, green squares, purple triangles). Other direct measurements are also included (pink square and black triangle). Results from this work from direct CTN measurements are shown as the red circles.
    \label{fig:LossAngle}}
\end{figure}

In order to compare our results with those published in the literature, we plot the coating loss angle as a function of temperature in Fig.~\ref{fig:LossAngle}.
The orange diamonds~\cite{Glasgow_PRL_2019}, green squares~\cite{Robie_thesis}, and blue triangles~\cite{Yamamoto_2006} are ringdown measurements that cannot distinguish between $\phi_{\perp}$ and $\phi_{\parallel}$.
The pink square~\cite{Matei_PRL}, black triangle~\cite{Chalermsongsak_2014} and purple diamond~\cite{Hafner_2020} are all from direct thermal noise measurements which assume $\phi_{\perp} = \phi_{\parallel}$.
For the sake of comparison, all values are plotted at 1 Hz.
We measured CTN at several temperatures near 4 K, and average these values together.
The x-axis error bar for that data point indicates the spread in temperature explored, while the y-axis error bar is the standard deviation of the extracted $\phi_c$.
We do not see a strong temperature dependence between 4 to 16 K.

We have presented direct measurements of the thermal noise for SiO$_2$/Ta$_2$O$_5$ HR coatings. These measurements are complementary to those attained by ringdown spectroscopy. For the design of future gravitational wave detectors, independent measurements of coating thermal noise are important. For ultrastable reference cavities, the use GaAs/AlGaAs crystalline coatings at cryogenic temperatures is important to explore. Other potentially low thermal noise optical coatings could also be tested using cryogenic ultrastable silicon cavities. 

We thank T. Bothwell, C. Kennedy and L. Sonderhouse for technical contributions.
We acknowledge funding support from DARPA, NIST, NSF QLCI Award 2016244, PTB, and by the Deutsche Forschungsgemeinschaft (DFG, German Research Foundation) under Germany’s Excellence
Strategy—EXC-2123 Quantum Frontiers.



\bibliography{sample}

\begin{thebibliography}{10}
\bibitem{FOOTNOTE}
Our measurements presented in this manuscript do not distinguish these two loss
  components. Without loss of generality we lump them together as a total loss
  angle. We also note a gradual shift away from the loss angle convention of
  $\phi_{\parallel}$ and $\phi_{\perp}$ in the LIGO community, and the loss
  angle can instead be defined in terms of bulk loss ($\phi_{B}$ and shear loss
  angles $\phi_{S}$~\cite{Rana_PRD}. Approximately, the assumption of $\phi_{B}
  = \phi_{S}$ is equivalent to that of $\phi_{\parallel} = \phi_{\perp}$,
  without detailed treatment of light penetration into the coating layers.
\end{thebibliography}

\end{document}